\documentclass[journal=jpclcd,manuscript=letter,linenumber]{achemso}
\usepackage{color}
\usepackage{soul}
\sethlcolor{yellow}
\usepackage{tabularx}
\usepackage{graphicx}
\usepackage{bm}
\usepackage{float}
\usepackage[version=3]{mhchem}

\newcommand{\expn}[1]{{\rm e}^{#1}}

\newcommand{\COMMENT}[1]{}
\newcommand{\textapprox}{{\raise.17ex\hbox{$\scriptstyle\mathtt{\sim}$}}}

\author{Shi Liu}
\affiliation{Geophysical Laboratory, Carnegie Institution for Science, Washington, D.C. 20015-1305 USA}
\email{sliu@carnegiescience.edu}
\author{Fan Zheng}
\affiliation{The Makineni Theoretical Laboratories, Department of Chemistry, University of Pennsylvania, Philadelphia, PA 19104--6323 USA}
\author{Ilya Grinberg}
\affiliation{The Makineni Theoretical Laboratories, Department of Chemistry, University of Pennsylvania, Philadelphia, PA 19104--6323 USA}
\author{Andrew M. Rappe}
\affiliation{The Makineni Theoretical Laboratories, Department of Chemistry, University of Pennsylvania, Philadelphia, PA 19104--6323 USA}
\email{rappe@sas.upenn.edu}

\title{Photoferroelectric and Photopiezoelectric Properties of Organometal Halide Perovskites}

\begin{document}

\abstract{
Piezoelectrics play a critical role in various applications. The permanent dipole associated with the molecular cations in organometal-halide perovskites (OMHPs) may lead to spontaneous polarization and thus piezoelectricity. Here, we explore the piezoelectric properties of OMHPs with density functional theory. We find that the piezoelectric coefficient depends sensitively on the molecular ordering, and that the experimentally observed light-enhanced piezoelectricity is due to to a non-polar to polar structural transition. By comparing OMHPs with different atomic substitutions in the $ABX_3$ architecture, we find that the displacement of the $B$--site cation contributes to nearly all the piezoelectric response, and that the competition between $A$--$X$ hydrogen bond and $B$--$X$ metal-halide bond in OMHPs controls the piezoelectric properties. These results highlight the potential of the OMHP architecture for designing new functional photoferroelectrics and photopiezoelectrics.\\
}

\newpage
\newpage
%\begin{linenumbers}
 
Organometal-halide perovskites (OMHPs) with their unprecedented rate of increasing power conversion efficiency (PCE)~\cite{KRICT} have transformed photovoltaic (PV) research and development. These materials, especially methylammonium lead iodide (MAPbI$_3$), are promising candidates for next-generation low cost-to-power PV technologies.~\cite{Kojima09p6050,Baikie13p5628,Im11p4088,Filip14p5757,Lee12p643,Etgar12p17396,Pogx87p6373,Stoumpos13p9019} Besides the solar cell application, the OMHP architecture, consisting of $ABX_{3}$ (organic monovalent cation, $A$; divalent metal, $B$; inorganic or organic anion, $X$), could serve as a promising platform for the design and optimization of materials with desired functionalities. By substituting $ABX_{3}$ atomic constituents, it is feasible to controllably tune the structural, electronic, optical, and magnetic properties of  functional materials.

Piezoelectrics (materials that can convert between electrical and mechanical energy) is a class of technological important materials with a variety of applications, ranging from waveguide devices, ultrasound transducers and fuel injectors to accelerometers and gyroscopes.~\cite{Jaffe12Book,Ballato95p916} The permanent dipole provided by the organic molecule in OMHPs is suggested to give rise to spontaneous polarization, which may play an important role in the photovoltaic working mechanism.~\cite{Frost14p2584,Liu15p693,Ma15p248} Switchable polar domains in $\beta$-MAPbI$_3$ have been observed via piezoresponse force microscopy (PFM).~\cite{Kutes14p3335} Though the robustness of the room-temperature ferroelectricity in MAPbI$_3$ is still under debate,~\cite{Xiao15p193,Fan15p1155,Kim15p1729,Coll15p1408} the piezoelectric response of MAPbI$_3$ has been demonstrated experimentally via PFM.~\cite{Coll15p1408} In particular, a five-fold increase of piezoelectric coefficient under illumination was observed.~\cite{Coll15p1408} Furthermore, the piezoelectric coefficient of MAPbI$_3$ under light ($\approx$25 pm/V) is comparable to inorganic ferroelectric thin films, such as lead zirconate titanate (PZT) thin films.~\cite{Lefki94p1764} The optically-tunable high piezoelectric response in MAPbI$_3$ suggests potential applications of OMHPs for optical piezoelectric transducers and light-driven electromechanical transistors. 

In this work, we investigate the piezoelectric properties of OMHPs with density functional theory (DFT). The interplay between the $A$-site molecular ordering and the macroscopic piezoelectric response is explored by quantifying the polarization-strain ($P$--$\eta$) and polarization-stress ($P$--$\sigma$) relationships in two model structures with designed molecular orientations. We further study the effect of atomic substitution with different organic molecules at the $A$ site (CH$_3$NH$_3^+$, CF$_3$NH$_3^+$), metal cations at the $B$ site (Pb$^{2+}$, Sn$^{2+}$, Ge$^{2+}$), and halide anions at the $X$ site (I$^-$, Cl$^-$). Our calculations show that the competition between the $A$--$X$ hydrogen bond and the $B$--$X$ metal-halide bond in OMHPs gives rise to a range of piezoelectric properties. The strong dependence of piezoelectric coefficients on the molecular ordering suggests that the light-enhanced piezoelectricity is likely due to the light-induced molecular reorientation under external voltage. This work demonstrates that the OMHP architecture could serve as a promising platform for designing functional piezoelectrics with robust optical tunability.

All calculations are performed with a plane-wave implementation of the Perdew-Burke-Ernzerhof (PBE) density functional~\cite{Perdew96p3865} in  the QUANTUM-ESPRESSO code.~\cite{Giannozzi09p395502} The optimized norm-conserving pseudopotentials~\cite{Rappe90p1227,Ramer99p12471} are generated using the OPIUM~\cite{Opium} code with the following electronic states included as valence states: Pb: 5$d$, 6$s$, 6$p$; I: 4$d$, 5$s$, 5$p$; C: 2$s$, 2$p$; N: 2$s$, 2$p$; H: 1$s$. A 4$\times$4$\times$4 Monkhorst-Pack grid is used, consistent with our previous work.~\cite{Zheng15p31,Liu15p693} The effect of molecular orientation on piezoelectricity is explored by calculating the piezoelectric coefficients of two model structures, namely M1 with non-zero total polarization and M2 with net-zero polarization, shown in Figure 1.  The M1 structure has an isotropic orientation of the MA cations in the $ab$-plane, and an anisotropic orientation with respect to the $c$-axis, giving rise to a net polarization along the $c$-axis. The molecular dipoles in the M2 structure adopt an antiferroelectric ordering and the structure is globally (nearly) paraelectric.  

First-principles techniques to determine piezoelectric coefficients have been widely applied to inorganic materials.~\cite{Saghi-Szabo99p12771,Saghi-Szabo98p4321,Bernardini02p4145,Bellaiche02p19} As piezoelectricity is a fundamental process incorporating different electromechanical effects, multiple piezoelectric equations exist.~\cite{Nye57book,Ballato95p916,Bellaiche02p19} The {\em direct} piezoelectric effect that links the induced polarization that develops in direction $i$ ($\Delta P_i$) with an applied stress with component $j$  ($\sigma_j$ in Voigt notation) can be described by 

\begin{equation}
\Delta P_i = d_{ij}\sigma_j
\end{equation}

\noindent where $\{d_{ij}\}$ is a third-rank tensor and each $d_{ij}$ parameter is usually referred to as a {\em direct} piezoelectric coefficient or piezoelectric strain coefficient, in units of pC/N. Another piezoelectric equation connecting the strain $\eta$ and the polarization is given by:

\begin{equation}
\Delta P_i = e_{ij}\eta_j
\end{equation}

\noindent where the $e_{ij}$ parameter is called the piezoelectric stress coefficient in the unit of C/m$^2$. It is noted that the $e_{ij}$ and $d_{ij}$ parameters are related to each other via elastic compliances and/or stiffness, though $d_{ij}$ is easier to measure experimentally. In this study, we focus on the evaluation of $d_{33}$ and $e_{33}$ (piezoelectric response along the $c$ axis). The total induced polarization along the $c$ axis can be expressed as $\Delta P_3 = e_{33}\eta_{3}+ e_{31}(\eta_{1} + \eta_2)$, where $\eta_{1} = (a-a_0)/a_0$, $\eta_{2} = (b-b_0)/b_0$ and $\eta_{3} = (c-c_0)/c_0$ are strains along the $a$, $b$, and $c$ axis, and $a_0$, $b_0$, and $c_0$ are lattice constants of an unstrained structure. When only applying strain along the $c$ axis, $e_{33}$ can be evaluated from the slope of polarization {\em vs} strain ($\eta_{3}$) curve. For a given strain $\eta_{3}$, the internal atomic coordinates are fully relaxed until the residual forces on atoms are less than 1.0$\times$10$^{-5}$ Ryd/Bohr, and the polarization of the optimized structure is determined with the Berry's phase approach. The estimation of $d_{33}$ with DFT is less straightforward, due to the difficulty of defining the stress tensor. We adopt the finite difference method developed in Ref~\citenum{Shi14p094105}. After applying a strain along the [001] direction, the other lattice parameters (lattice constants $a$ and $b$, and lattice angles) as well as internal atomic coordinates are fully optimized until all stress tensor elements except $\sigma_3$ are smaller than 1.0$\times$10$^{-3}$ GPa. The polarization values of various strained structures are then calculated, and the slope of $P$ {\em vs} $\sigma_3$ determines the value of $d_{33}$. Throughout the calculations, $\pm 0.5\%$ strains are applied to the fully optimized ground-state structures to obtain linear response. 

A previous first-principles benchmark study highlighted the importance of dispersive interactions and the plane-wave cutoff energy ($E_c$) in accurately determining structural properties of OMHPs.~\cite{Egger14p2728} Because the value of piezoelectric coefficient depends sensitively on the lattice constants, we first compare the optimized structural parameters of MAPbI$_3$ obtained with four different methods: pure PBE density functional with $E_c=50$~Ry, PBE with $E_c=80$~Ry, PBE plus Grimme dispersion correction~\cite{Grimme06p1787}(PBE-D2) with $E_c=50$~Ry, and PBE-D2 with $E_c=80$~Ry. As shown in Table 1, the value of unit cell volume predicted by PBE-D2 with $E_c=80$~Ry best agrees with the experimental results. Therefore, we choose this method for the following piezoelectric coefficient calculations.

Figure 2 shows the $P$-$\sigma_3$ and $P$-$\eta_3$ curves for MAPbI$_3$ in polar M1 and non-polar M2 configurations. The computed piezoelectric coefficients of M1 are $d_{33}$ = 31.4 pm/V and $e_{33}$ = 0.83 C/m$^2$, dramatically higher than those of M2,  $d_{33}$ = 3.1 pm/V and $e_{33}$ = 0.07 C/m$^2$. Note that the values of $d_{33}$ of M1 and M2 agree reasonably well with experimental values~\cite{Coll15p1408} measured under white light (25 pm/V) and dark (6 pm/V), respectively.  It is suggested that illumination could weaken the hydrogen bond between the MA cation and the inorganic Pb-I scaffold, thus effectively reducing the rotational barrier of the MA cation.~\cite{Gottesman14p2662} This allows the formation of a more polar structure with aligned MA cations in the presence of an electric field, which could arise from the photovoltage across the thin film. Therefore, the light-enhanced piezoelectric response is likely due to the transformation of a configuration with less-ordered molecular orientation ({\em e.g.}, M2--like) to a configuration with more-ordered molecular orientation ({\em e.g.}, M1--like).

To reveal the origin of piezoelectricity, we decompose the total polarization into the contributions from the $A$-site MA cations and $B$-site Pb atoms in M1. The polarization from Pb is estimated with $P_{\rm Pb}$ = $Z_{33}^*({\rm Pb}) \times D({\rm Pb})/V_{\rm u}$, where $V_{\rm u}$ is the volume of the primitive unit cell (1/4 of the volume of the tetragonal supercell), $Z_{33}^*({\rm Pb})$ is the Born effective charge of Pb and $D({\rm Pb})$ is the average displacement of Pb along the $c$ axis with respect to the center of its I$_6$ cage. The calculated value of $Z_{33}^*({\rm Pb})$ is +4.24, significantly larger than the nominal charge of Pb (+2.0) in a pure ionic picture. This indicates the strong covalency for the Pb--I bonds and the presence of dynamic charge transfer coupled with the change of Pb-I bond length. The polarization from MA$^+$ molecules is approximated with $P_{\rm MA}= \mu\cos (\theta) / V_{\rm u}$, where $\mathbf{\mu }$ is the dipole of MA$^+$ molecule and $\theta$ is the average angle between the molecular dipole and the $c$ axis. The site-resolved polarization and $d_{33}$ are presented in Figure 3a. It is clear that though both Pb displacements and molecular dipoles are responsible for the total polarization, Pb atoms contribute nearly all the piezoelectric response, with the contribution from the MA$^+$ molecules negligible. As shown in Figure 3b, the angle between the MA$^+$ and the $c$ axis remains nearly constant, while the atomic displacement of Pb changes linearly with the stress.

We further explore the effect of chemical substitution on the piezoelectric properties of OMHPs. Table 2 presents the values of $d_{33}$ and $e_{33}$ for MAPbI$_3$, MASnI$_3$, MAGeI$_3$, CF$_3$NH$_3$PbI$_3$, MAPbCl$_3$ and MASnCl$_3$. For all the materials investigated, the M2 configuration has lower polarization and smaller $d_{33} $ and $e_{33}$ than the M1 configuration. It is noted, however, that MAGeI$_3$ of M2 configuration still has a moderate polarization of 8.92$\mu$C/cm$^2$ despite the molecular dipoles nearly canceling each other out. Close examination of the structure reveals that the Ge atom is displaced by 0.23 \AA~away from the center of GeI$_6$ cage. The significant displacement is due to the small size of Ge relative to the large $A$ cations and I$_6$ octahedron. As shown in Figure 4, for MAPbI$_3$, MASnI$_3$ and MAGeI$_3$, three OMHPs differing only by the $B$-site element, MAGeI$_3$ M1 has the highest polarization, while MASnI$_3$ M1 has the highest piezoelectric coefficient ($d_{33}$ = 100.9 pC/N), comparable to classic inorganic ferroelectric PbTiO$_3$ ($d_{33}$ = 80 pC/N).~\cite{Shi14p094105} The higher $d_{33}$ in MASnI$_3$ is attributed to the larger Born effective charge $Z_{33}^*({\rm Sn})=+4.64$, compared to $Z_{33}^*({\rm Ge})= +3.60$ and $Z_{33}^*({\rm Pb})=+4.24$. Comparing MAPbCl$_3$ and MASnCl$_3$ to MAPbI$_3$ and MASnI$_3$, we find that substituting I with Cl dramatically reduces $d_{33}$ and $e_{33}$, which can be explained by the reduced covalency (more ionic character) of Pb(Sn)-Cl bonds. Born effective charge calculations show that $Z_{33}^*$ of Sn is reduced to +3.39 in MASnCl$_3$. This means a smaller charge transfer, thus less polarization change, will occur upon change of the Sn-Cl bond length, causing smaller piezoelectric response.\\

As demonstrated with CF$_3$NH$_3$PbI$_3$, replacing the MA$^+$ with a more polar molecular cation greatly increases both the polarization and the piezoelectric response. The larger molecular dipoles not only contribute directly to the polarization, but also increase the hydrogen bonding strength between NH$_3$ and I, which indirectly weakens the bonding between I and Pb. This is supported by the larger Pb displacement ($\approx 0.30$ \AA) at zero stress condition compared to that of $\approx 0.09$ \AA~in MAPbI$_3$. In other words, the Pb atoms in CF$_3$NH$_3$PbI$_3$ have a softer potential energy surface and therefore a stronger response to external perturbation, resulting in higher piezoelectricity. 

Here, we discuss other possible mechanism for the light-enhanced piezoelectricity in more detail. It is suggested that the light illumination, by weakening the hydrogen bond between the MA cation and the inorganic PbI$_3$ sublattice, could make the molecular cations easier to rotate and to align with external field. This is supported by DFT calculations which reveal a reduced binding energy of the MA cation to the inorganic cage at the triplet state.~\cite{Gottesman14p2662} Additionally, as the band gap of MAPbI$_3$ is in the visible light region, a significant amount of above band gap of excitation occurs with the excess energy dissipating as heat (kinetic energy of molecules) when the hot carriers relax to the band edges. We have recently shown that the molecules play an important role in carrier relaxation, with MA translation, CH/NH twisting, and CH/NH stretching modes  
assisting the process.~\cite{Zheng15Nano} We thus develop a simple analytical model here to examine the magnitude of light-induced thermalization in a MAPbI$_3$ thin film. Shown in Figure 5a is the absorption spectrum $\alpha(E)$ computed from the DFT optical dielectric function. The dielectric function is summed over the Brillouin zone and enough empty bands to converge. Assuming that all the energy released from hot carrier relaxation is transformed into kinetic energy ($\kappa$) of molecules, the average thermal energy received by each molecule at depth $D$ per second can be estimated as:

 \begin{equation}
  \kappa\left(D\right)  = \frac{1}{A}\int_{\rm E_{\rm g}}^{\infty} dE \frac{I_0\left(E\right)\expn{-\alpha\left(E\right)D}\left(E-E_{\rm g}\right)}{E}\left(1-\expn{-\alpha\left(E\right)\Delta d}\right)
 \end{equation}
 
\noindent where $I_0(E)$ is the sunlight spectrum, $\Delta d$ is the thickness of  a single layer of molecules (equal to the lattice constant of MAPbI$_3$, 0.63 nm), $A$ is the number of molecules per unit area, and $E_{\rm g}$ is the band gap of 1.65 eV calculated from DFT. As plotted in Figure 5b, the energy obtained by a single molecule per second has a strong dependence on the depth due to the exponential decay of the light intensity. In particular, the molecules near the top surface of the sample can obtain significant energy (much larger than room temperature energy scale) to overcome the barrier and re-orient their dipole directions under the applied voltage, which is likely to result in a more polar structure with large piezoelectricity, as measured in the experiments. 

In summary, the piezoelectric properties of several OMHPs are quantified with {\em ab initio} density functional theory. For MAPbI$_3$, the calculated piezoelectric coefficients of the polar and non-polar configurations agree reasonably well with experimental values measured under white light and dark, respectively. This suggests a light-driven molecular reordering, 
which could be attributed to the weakened hydrogen bonds between MA cations and inorganic cages in the excited state and the thermalization arising from hot carrier relaxation. The photopiezoelectricity of OMHPs offers a potential avenue to optical-transducers. Comparing the piezoelectric properties of OMHPs with different organic molecules, metal cations and halide anions helps to identify several design principles for tuning the piezoelectric coefficients. We find that the atomic displacement of $B$--site metal cation is responsible for the piezoelectric response, and therefore creating a softer energy profile {\em vs} cation displacement is critical for enhancing $d_{33}$ and $e_{33}$. Our finding highlights the potential of OMHPs for the design and optimization of functional photopiezoelectrics and photoferroelectrics. 
%\end{linenumbers}

\newpage
\begin{table*}[!]
  \centering 
    \caption{Structural parameters of tetragonal MAPbI$_3$ in the M1 configuration, compared to experimental data.}
    \resizebox{1.0\textwidth}{!}
  {
  \begin{tabular}{cccccccccc}
  \hline 
  \hline
  & Method &$E_c$ (Ry)&$a$ (\AA)&$b$ (\AA)&$c$ (\AA)&$\alpha$&$\beta$&$\gamma$&$V$(\AA$^3$)\cr
  \hline
M1&PBE&50& 8.91 &8.81 &13.09 &90.00&90.00&90.03&1027.10\cr
     &PBE&80& 8.97 &8.92 &13.29 &90.01&90.03&90.00&1062.87\cr
      &PBE-D2&50& 8.64 &8.68 &12.77 &90.05&89.86&89.92&957.93\cr
     &PBE-D2&80& 8.67 &8.74 &12.97 &90.07&89.87&90.00&981.90\cr
  \hline
M2&PBE&50& 8.88 &8.89 &13.06 &91.18&91.39&89.75&1030.33\cr
     &PBE&80& 8.99 &8.96 &13.22 &91.27&90.04&89.71&1063.54\cr     
     &PBE-D2&50& 8.63 &8.64 &12.82 &90.41&90.45&90.17&955.71\cr
     &PBE-D2&80& 8.69 &8.69 &12.97 &90.09&90.10&90.06&979.83\cr
  \hline
 &experiment& Ref \citenum{Stoumpos13p9019}& 8.85 & 8.85 & 12.64 &90.0 &90.0 & 90.0 & 990.00\cr 
 &          & Ref~\citenum{Pogx87p6373}& 8.86& 8.86 & 12.66 &90.0 &90.0 & 90.0 & 993.80\cr 
 &          & Ref~\citenum{Baikie13p5628}  & 8.85 & 8.85 & 12.44 &90.0 &90.0 & 90.0& 974.33\cr 
  \hline
  \hline
  \end{tabular}
  } %resizebox 
\end{table*}

\newpage
\begin{table*}[!]
  \centering 
  \caption{Ground state polarization along the $c$ axis ($P_c$) in $\mu$C/cm$^2$, piezoelectric coefficients $d_{33}$ in pC/N and $e_{33}$ in C/m$^2$ for various OMHPs of M1 and M2 configurations}
  \resizebox{0.7\textwidth}{!}
  {
  \begin{tabular}{l|ccc|ccc}
  \hline 
  \hline
 % & \multicolumn{3}{c}{M1}& \multicolumn{3}{c}{M2} \cr
 & &M1&&&M2&\cr
  \hline
  OMHP& $ P_c$ &$d_{33}$ & $e_{33}$&$P_c$ & $d_{33}$ & $e_{33}$\cr
  \hline
 MAPbI$_3$   &5.59 &31.4   &0.83    &0.07&3.1&0.07\cr
  MASnI$_3$  &7.75 &100.9  &1.85 &-0.22&23.7&0.51\cr
 MAGeI$_3$   &13.68 &27.4  &0.49  &8.92&32.4&0.52\cr
 MAPbCl$_3$ &5.96   &6.7  &0.16    &-0.01&--&--\cr
 MASnCl$_3$ &5.46  &4.1 & 0.05      &0.01&--&--\cr
 CF$_3$NH$_3$PbI$_3$&15.8&248 &1.59&-0.09&4.23&0.16\cr
  \hline
  \hline
  \end{tabular}
  } %resizebox 
\end{table*}
%\newpage

\newpage
%Ref
%\bibliography{Piezo.bib}
\providecommand{\latin}[1]{#1}
\providecommand*\mcitethebibliography{\thebibliography}
\csname @ifundefined\endcsname{endmcitethebibliography}
  {\let\endmcitethebibliography\endthebibliography}{}

\newpage
%%Figures
\begin{figure}
\centering
\includegraphics[width=1\columnwidth]{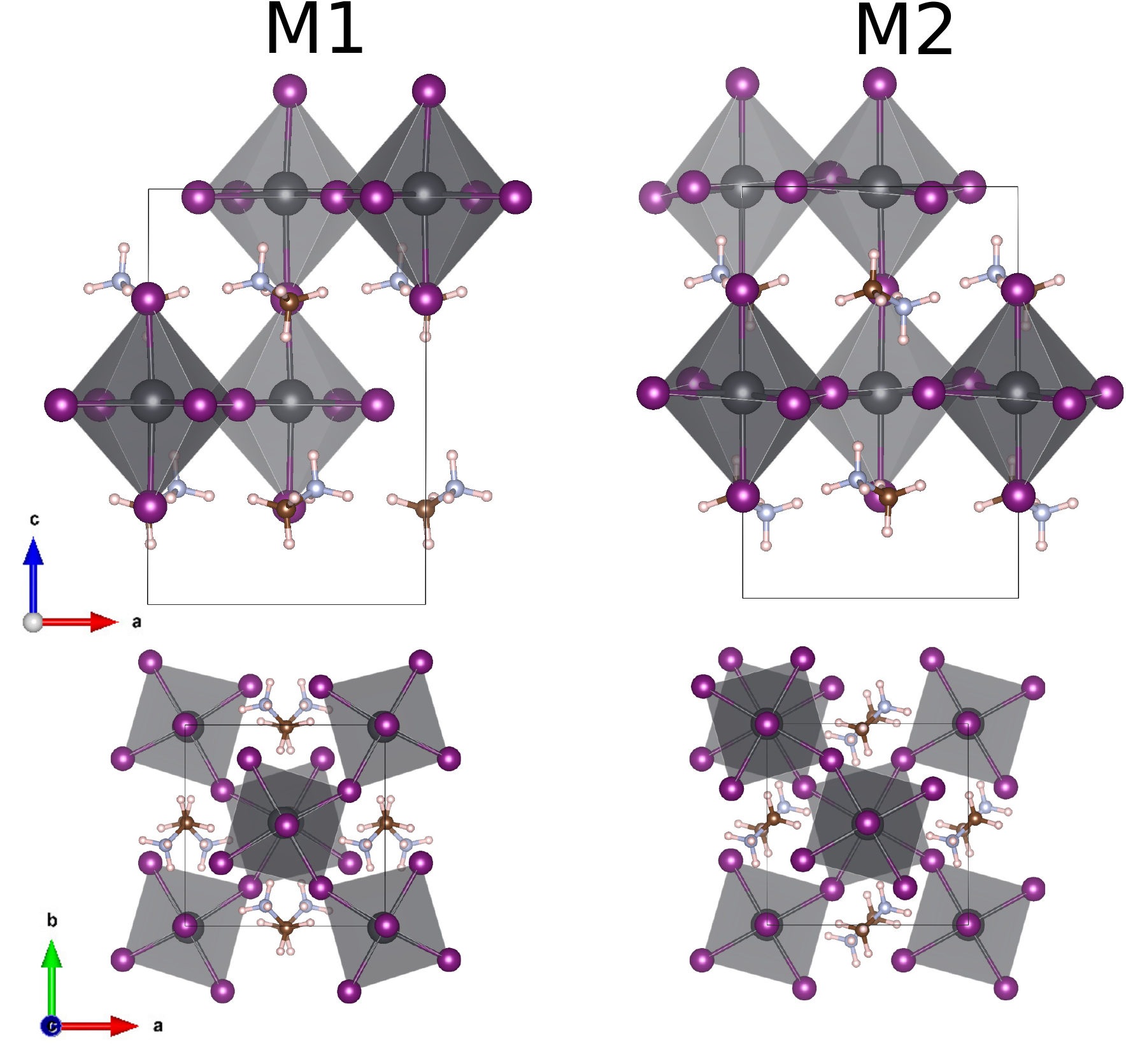}\\
 \caption{Side view and top view of polar (M1) and non-polar (M2) structural models for MAPbI$_3$. The M1 configuration has molecular dipoles aligned preferentially along $c$ axis while distributed isotropically within the $ab$ plane. The molecules in the M2 configuration adopt an antiferroelectric arrangement.} 
 \end{figure}
\newpage

\begin{figure}
\centering
\includegraphics[width=1\columnwidth]{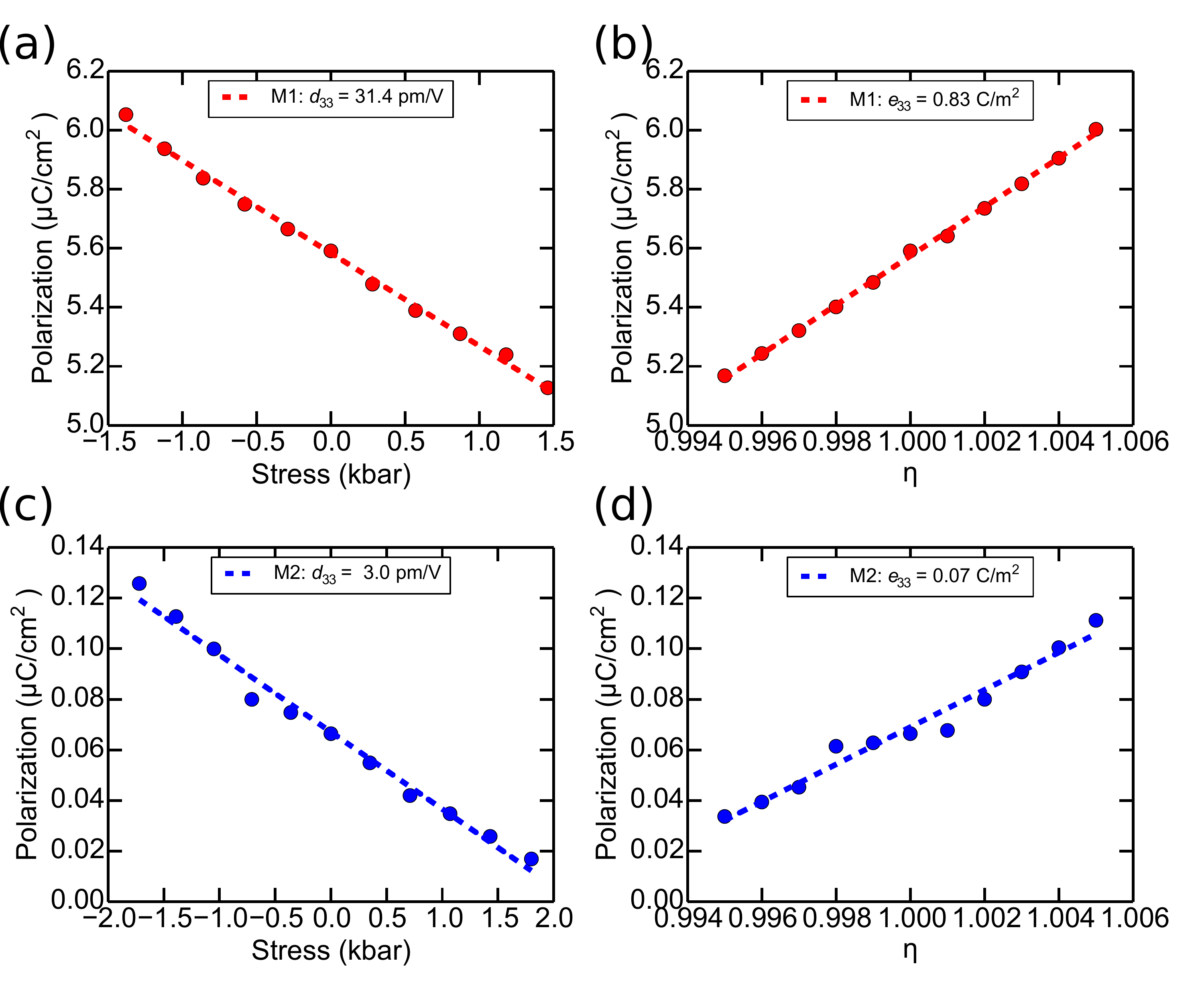}\\
 \caption{Piezoelectric properties of MAPbI$_3$ in M1 (a-b) and M2 (c-d) configurations. (a,c) Polarization as a function of stress. (b,d) Polarization as a function of strain.}
 \end{figure}
\newpage

\begin{figure}
\centering
\includegraphics[width=1\columnwidth]{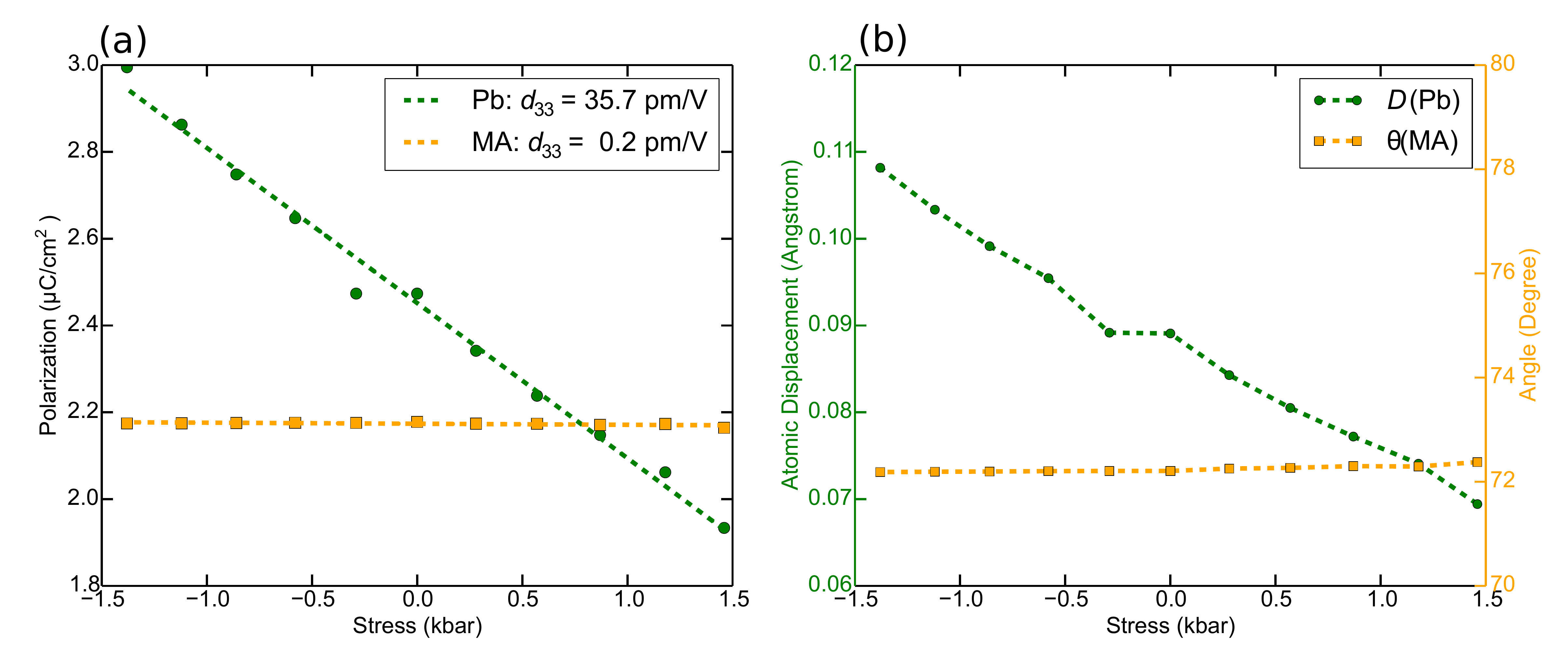}\\
 \caption{(a) Site-resolved polarization as a function of stress. (b) Atomic displacement and molecular orientational angle as a function of stress.}
 \end{figure}
\newpage

\begin{figure}
\centering
\includegraphics[width=1\columnwidth]{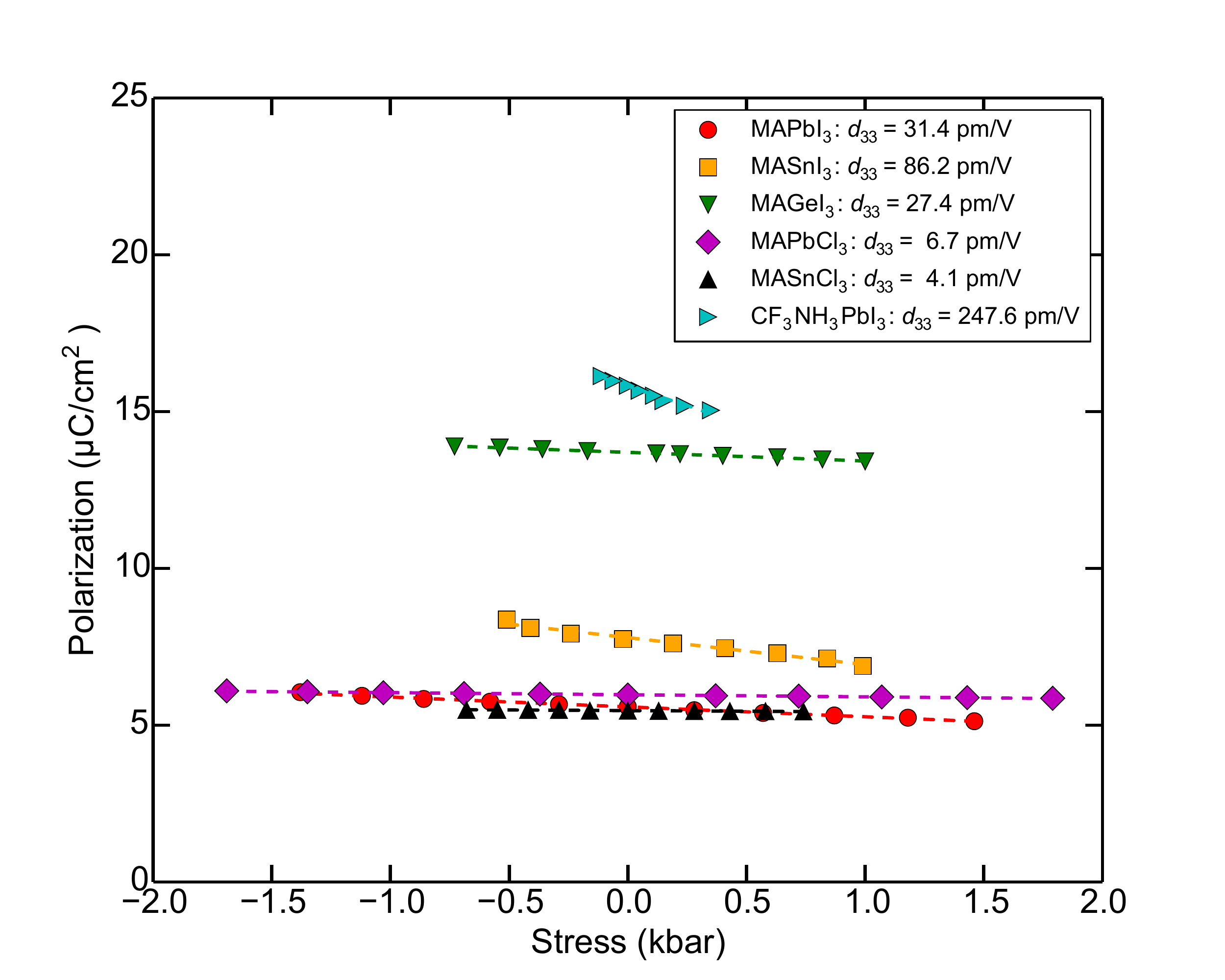}\\
 \caption{Stress-dependence of polarization for various OMHPs, all with M1 molecular orientation.}
 \end{figure}
 \newpage
 
\begin{figure}
\centering
\includegraphics[width=1\columnwidth]{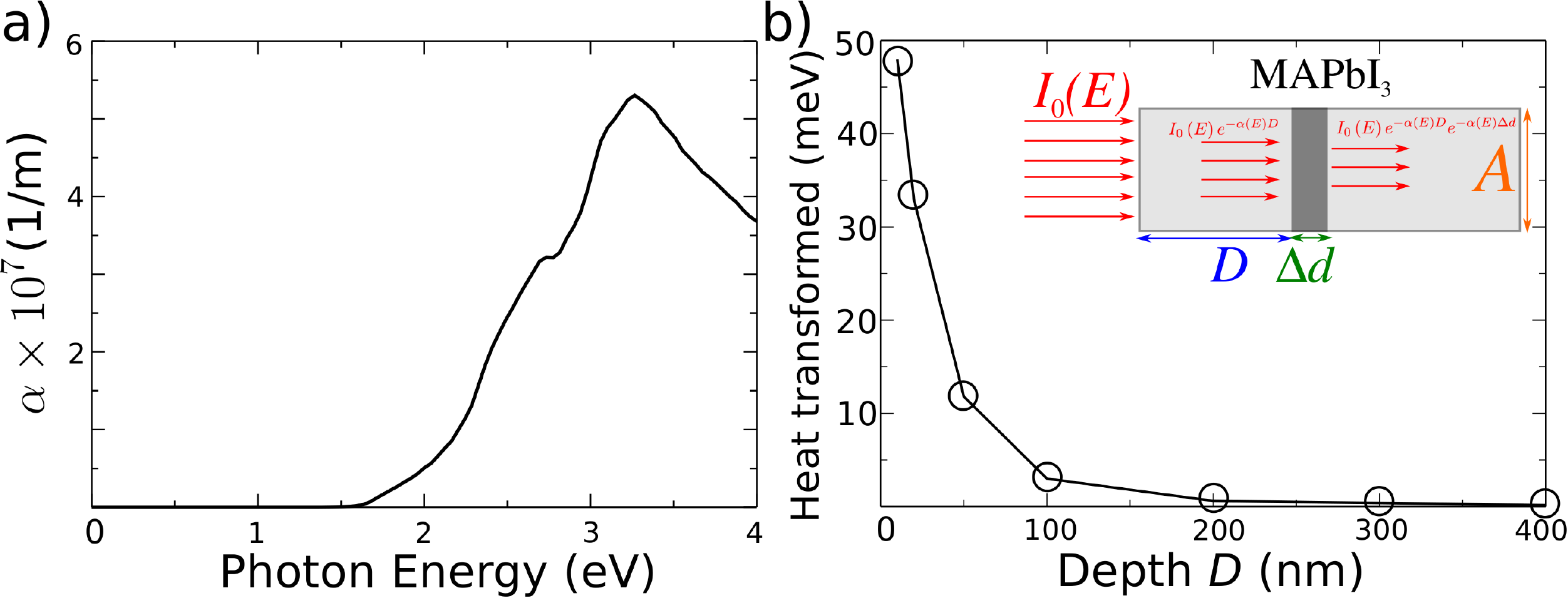}\\
 \caption{(a) The DFT calculated absorption spectrum. (b) Estimation of the averaged energy transferred from the photo-excited carriers to each molecule at different depths ($D$) per second. The insert illustrates the model described by Equation 3.  }
 \end{figure} 
 \newpage

\noindent{\bf {\Large Graphical TOC Entry}}
\begin{figure}[H]
\centering
\includegraphics[width=12cm, clip]{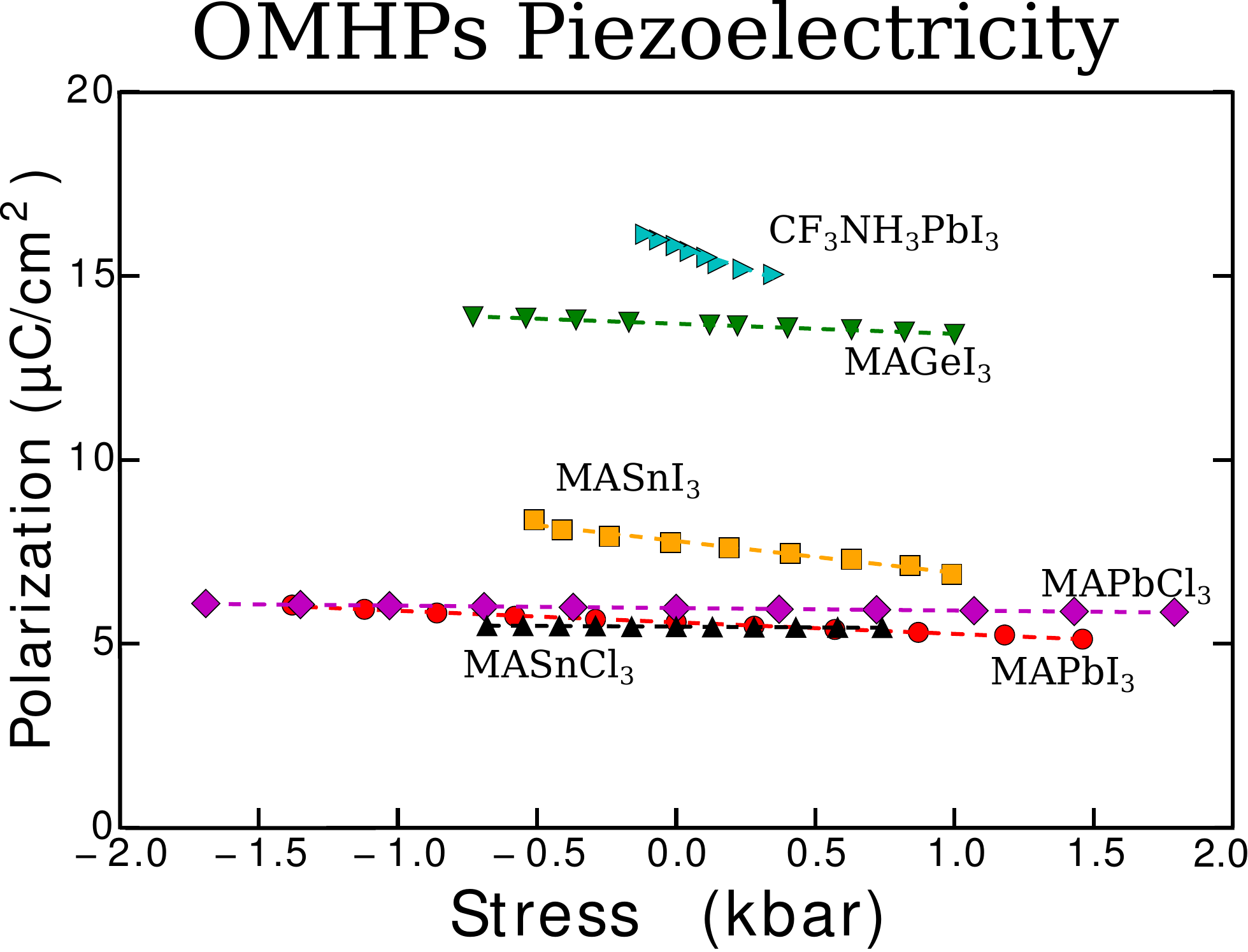}
\end{figure}

\end{document}